\documentclass[journal]{IEEEtran}

\usepackage{cite,graphicx,amsmath,array,subfigure,url,booktabs,xcolor,multirow,color,balance}

\begin{document}
\title{Auxiliary Factor Method to Remove ISI of Nyquist Filters}
\author{Zijian Zhou, \IEEEmembership{Member,~IEEE}, Lifeng Lin, and Bingli Jiao, \IEEEmembership{Senior Member,~IEEE}
\thanks{This work was jointly supported by the National Key Research and Development Program of China under Grants 2018YFB2202202, the National Natural Science Foundation of China under Grant 61961016 and 62171006, and the National Key Laboratory Foundation 2021-JCJQ-LB-006 under Grant 6142411222113. The calculations were supported by the High-Performance Computing Platform of Peking University. \emph{(Corresponding author: Bingli Jiao.)}}
\thanks{Z. Zhou, L. Lin, and B. Jiao are with the School of Electronics, Peking University, Beijing 100871, China (e-mail: zjzhou1008@pku.edu.cn, linlifeng@pku.edu.cn, jiaobl@pku.edu.cn).}}
\maketitle

\begin{abstract}
As has been known, the Nyquist first condition promises no intersymbol interference (ISI) as derived in the frequency domain.  However, the practical implementation using the FIR filter truncates the Fourier transform by its window and prevents the mathematical calculation from reaching the ideal solution at zero-ISI.  For obtaining better results, an increase in the window's length is required in general.  To address this problem, a new approach is presented by using auxiliary factors (AFs) to compensate shortcomings of the truncated Fourier transform and remove the ISI completely, regardless of the window's length.  In addition, the performance in the presence of the timing jitter is also improved significantly.  The closed-form solution of the AFs is derived and the effectiveness is confirmed by the simulation results.  Finally, the problems of the transmission delay and additional calculation complexity are analysed.
\end{abstract}

\begin{IEEEkeywords}
Nyquist filter, intersymbol interference (ISI), auxiliary factor (AF), timing jitter, pulse shaping.
\end{IEEEkeywords}

\section{Introduction}
\IEEEPARstart{T}{he} baseband pulse shaping signal can be transmitted within the limited bandwidth without any intersymbol interference (ISI) in principle, when the first Nyquist condition is satisfied \cite{ref_Nyquist, ref_Proakis}.  In most applications, the implementation uses the finite impulse response (FIR) filter for the convenience of the transformation to the time domain.  However, since the window of the FIR filter truncates the Fourier transform at a finite length, the mathematical minimisation process cannot annihilate the calculation error and, thus, gives a rise to the residual ISI.

To pursue a better approximation, the previous studies focus their attention on the following two types of approaches.  One is the design of the profile functions of the Nyquist frequency response.  A family of the solutions has been obtained, e.g., ``better than raised cosine'' \cite{ref_Beaulieu}, farcsech \cite{ref_Assalini}, conjugate-root \cite{ref_Tan}, and polynomial \cite{ref_Chandan}.

The other proposes new objective functions and uses the iterative algorithm to optimise the FIR filter. These methods include the optimum asymmetric filter \cite{ref_Chevillat} to maximise the spectral energy in the passband, the Remez-type procedure of two-stage FIR Nyquist filter \cite{ref_Saramaki} to optimise the subfilters, the equiripple method \cite{ref_Boroujeny} to minimise the energy in the tails, and the novel frequency mask Nyquist filter \cite{ref_Traverso} to realise a high stop-band attenuation.

Although the aforementioned methods show improvements one another, the theoretical imperfectability remains still due to the truncation of the Fourier expansion.

This work presents a method to remove the ISI completely by introducing the auxiliary method to compensate shortcomings of the truncated Fourier expansion \cite{ref_Jiao}. The proposed method, referred to as the auxiliary factor (AF) method, combines the conventional Nyquist filters to get rid of the dependency on the window's length.

For adapting the application to most communication systems, the AF method is formulated to work with the paired Nyquist filters: One filter is located at the transmitter for limiting the transmission bandwidth and the other at the receiver for maximising the signal-to-noise (SNR) ratio.  In the AF method, the perfect zero-ISI performance can be achieved at the various roll-off factors. Also, the low sensitivity to the timing jitter is found and the nullification of the signal distortion at the boundary is realised \cite{ref_Zhou}.

The rest of the paper is organised as follows. Section II introduces the system model and the signal formulations, and Section III derives the closed-form solution. The performance evaluation is provided in Section IV and the paper is concluded in Section V.

\emph{Notation}: Boldfaced lowercase letter represents a vector, e.g., ${\mathbf{a}}=[a_0,a_1,\cdots,a_k,\cdots,a_K]$, where $a_k$ is the $k$\textsuperscript{th} component, and boldfaced capital letter denotes a matrix, e.g., $\mathbf{A}$, where $a_{ij}$ is the element in the $i$\textsuperscript{th} row and the $j$\textsuperscript{th} column.  Superscript $\left( \cdot \right)^{\dag}$ and $\left( \cdot \right)^{-1}$ stand for transposition and inversion operation, respectively.

\section {System Model and Signal Formulation}
\begin{figure}
\centering
\vspace{-0.15cm}
\includegraphics[width=0.85\columnwidth]{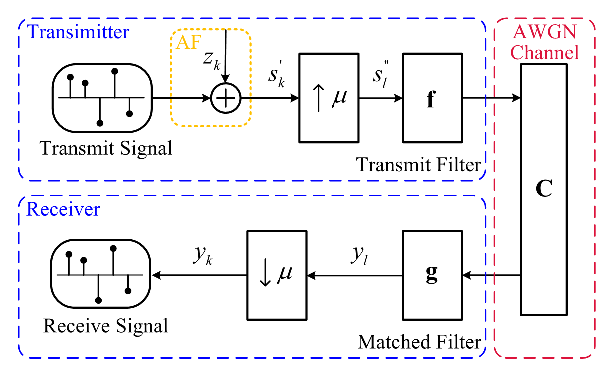}
\caption{The structure of the paired Nyquist filters working with the AF method.}
\vspace{-0.2cm}
\label{fig_1}
\end{figure}

Consider the communication system as shown in Fig. \ref{fig_1}.  The signs $ \uparrow\mu$ and $\downarrow\mu$ denote the upsampling and downsampling operators with the ratio $\mu$, respectively.  As the same as most studies on Nyquist filters, the additive white Gaussian noise (AWGN) channel model is adopted for the signal transmission between the transmitter and receiver.

Let us assume that the transmit signal block, i.e., $\mathbf{s} = \left[s_0,s_1,\cdots,s_k,\cdots,s_K\right]$, and the paired Nyquist filters' coefficients, i.e., $\mathbf{f} = \left[f_{-\frac{M}{2}},f_{-\frac{M}{2}+1},\cdots,f_m,\cdots,f_{\frac{M}{2}}\right]$ and $\mathbf{g} = \left[g_{-\frac{M}{2}},g_{-\frac{M}{2}+1},\cdots,g_n,\cdots,g_{\frac{M}{2}}\right]$, where $K+1$ is the block length and the even integer $M$ is the filters' order, are perfectly known at the transmitter.  The AF set, i.e., $\mathbf{z} = \left[z_0,z_1,\cdots,z_k,\cdots,z_K\right]$, is added to the transmit signal block for removing the ISI completely written as
\begin{equation} \label {2-1}
\mathbf{s}' = \mathbf{s} + \mathbf{z},
\end{equation}
where $\mathbf{s}'=[s'_0,s'_1,\cdots,s'_k,\cdots,s'_K]$ denotes the compensated signal block with $s'_k = s_k+z_k$.

The upsampling operation is carried out by interpolating $\mu-1$ zeros uniformly between two components of the compensated signals, e.g., between $s'_k$ and $s'_{k+1}$.  The upsampled signal can be expressed as 
\begin{equation} \label {2-2}
s''_l = \left( {\left\lfloor {\frac{l}{\mu}} \right\rfloor  - \left\lfloor {\frac{l - 1}{\mu}} \right\rfloor } \right)
s'_{\left\lfloor \frac{l}{\mu } \right\rfloor}  \ \ {\text{for}} \ \ l=0,1,\cdots,L,
\end{equation}
where $s''_l$ is the $l$\textsuperscript{th} upsampled signal with $L = \mu (K+1) - 1$ and $\left\lfloor{\cdot}\right\rfloor$ represents the floor function.

Passing $s''_l$ through the transmit filter yields
\begin{equation} \label {2-3}
x_l = \sum\limits^{\frac{M}{2}}_{m = -\frac{M}{2}} {{s''_{l - m}} {f_m}},
\end{equation}
where $x_l$ is the $l$\textsuperscript{th} output of the transmit filter.  At the receiver, $x_l$ is taken to the input of the matched filter and the output of the filter is obtained by
\begin{equation} \label {2-4}
\begin{split}
y_l = \sum\limits^{\frac{M}{2}}_{m = -\frac{M}{2}} \sum\limits^{\frac{M}{2}}_{n = -\frac{M}{2}} {{s''_{l - m - n} {f_m} {g_n}}}.
\end{split}
\end{equation}
It is noted that the noise term is omitted in \eqref{2-4} as the same in most theoretical studies on the filter design.

The zero-ISI can be achieved by solving
\begin{equation} \label {2-5}
y_k - s_k = 0~\text{for}~k = \left\lfloor \frac{l}{\mu} \right\rfloor = 0,1,\cdots,K
\end{equation}
before the signal transmission.

\section {Solution of The AF Method}
To find the solution of the AF method,  \eqref{2-5} is expressed explicitly in terms of $z_k$ as
\begin{equation} \label {3-1}
\sum\limits^{\frac{M}{2}}_{m = -\frac{M}{2}} {\sum\limits^{\frac{M}{2}}_{n=-\frac{M}{2}} {\left(s_{\left\lfloor k - \frac{m+n}{\mu} \right\rfloor} + z_{\left\lfloor k - \frac{m+n}{\mu} \right\rfloor} \right) {f_m}{g_n}} }  - {s_k} = 0
\end{equation}
for $k=0,1,\cdots,K$.

Since the index $l$ of $s''_l$ and $x_l$ must be a non-negative integer and not larger than $L$, the following constraints of the indexes $m$, $n$, and $l$ in \eqref {2-3} and \eqref {2-4} are required by
\begin{equation} \label {3-2}
\left\{ \begin{split}
0 &\le l - m \le L\\
0 &\le l - n \le L\\
0 &\le l-m-n \le L.
\end{split} \right.
\end{equation}
In addition, for a given $k$ in \eqref{3-1}, the index $m+n$ should be restricted by 
\begin{equation} \label {3-3}
\left| {m + n} \right| \le M.
\end{equation}

To facilitate our derivations, the constraint \eqref{3-2} and \eqref{3-3} for $m$ and $n$ can be changed to a constraint using $p$ by
\begin{equation} \label {3-4}
\left\{ \begin{split}
-\frac{M}{\mu } &\le p \le \frac{M}{\mu} \\
k - K &\le p \le k,
\end{split} \right.
\end{equation}
where $\mu p = m+n$ is assumed.  The constraints of $m$ and $n$ for $-\frac{M}{2}\le m \le \frac{M}{2}$ and $-\frac{M}{2}\le n \le \frac{M}{2}$ can be changed to the following constraint using $p$ and $n$ as
\begin {equation} \label {3-5}
\left\{ \begin{split}
-\frac{M}{2} &\le m \le \frac{M}{2} \\
\mu p -\frac{M}{2} &\le m \le \mu p + \frac{M}{2} \\
\mu k- \mu \left( K+1 \right) + 1 &\le m \le \mu k.
\end{split} \right.
\end {equation}

Based on the above two constraints, the upper and lower bounds of the summation in \eqref {3-1} can be replaced by
\begin{equation} \label {3-6}
\sum\limits^{P^{(2)}_k}_{p=P^{(1)}_k} {\left( {{s_{k - p}} + {z_{k - p}}} \right)\sum\limits^{M^{(2)}_k}_{m =M^{(1)}_k} {{f_{\mu p - m}}{g_m}} }  - {s_k} = 0,
\end{equation}
where $P^{(2)}_k$ and $M^{(2)}_k$ are the upper bounds and $P^{(1)}_k$ and $M^{(1)}_k$ are the lower bounds in the condition of \eqref {3-4} and \eqref {3-5}. The four bounds of the summation can be expressed as
\begin{equation} \label {3-7}
\left\{ \begin {split}
P^{(1)}_k&={\rm{max}}\left\{ -2\tau_0,k-K \right\}\\
P^{(2)}_k&={\rm{min}}\left\{ 2\tau_0,k \right\}\\
M^{(1)}_k&={\rm{max}}\left\{-\mu \tau_0,\mu \left( p-\tau_0 \right),\mu \left(k - K - 1 \right) + 1 \right\}\\
M^{(2)}_k&={\rm{min}}\left\{\mu \tau_0,\mu \left( p+\tau_0 \right), \mu k \right\}
\end {split} \right.
\end{equation}
where $\tau_0 = \frac{M}{\mu}$ is the number of sidelobes of the  AF-based filters.  The head and terminal distortions at the output of the matched filter are also removed in the AF method because the boundary effect has been considered in \eqref {3-4} and \eqref {3-5}.

\begin{figure}
\centering
\vspace{-0.35cm}
\subfigure[Head distortion]{
\centering
\includegraphics[width=0.95\columnwidth]{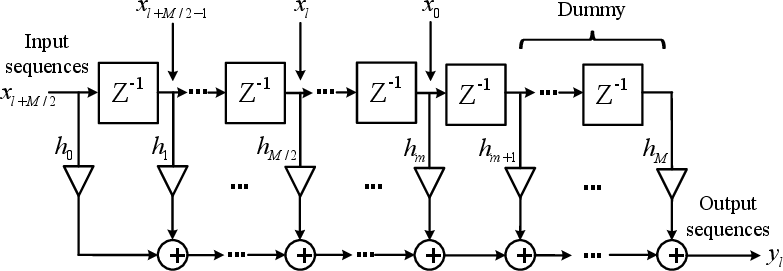}
}
\subfigure[Terminal distortion]{
\centering
\includegraphics[width=0.95\columnwidth]{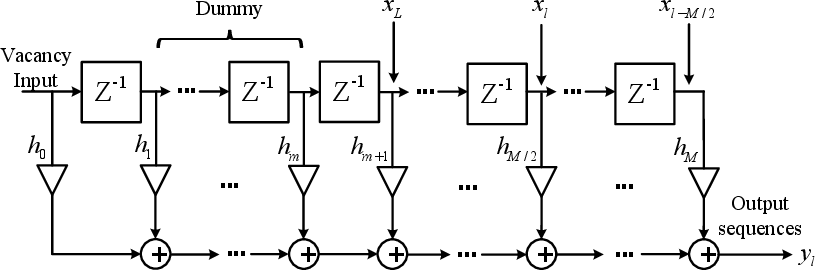}
}
\caption{Diagrams of the boundary effect: (a) and (b) illustrate the examples of head and terminal problems, respectively.}
\label{fig_2}
\end{figure}

It is noted that the boundary effect of the FIR filter was studied in \cite {ref_Zhou}.  Here, a short explanation is given to the head and terminal terms as follows.  As has been known, in the operation of an FIR filter, the signal samples are multiplied by the weights one another when they go from the input to the output of the filter. Figure \ref{fig_2}(a) shows an example of the head problem when the first input signal sample, i.e. $x_0$, arrives at weight $h_m$ before the last weight $h_M$.  Then, the weights from $h_{m+1}$ to $h_M$ are actually the dummy weights that do not work appropriately according to the FIR filter's theory.  While, Figure \ref{fig_2}(b) plots the terminal problem when the last input signal sample arrives at $h_{m+1}$.  In this case, the weights from $h_0$ to $h_m$ become the dummies.  The head and terminal parts cause the boundary effect mentioned above.

To proceed with the derivations, we express \eqref{3-6} in form of a matrix as 
\begin{equation} \label {3-8}
\mathbf{A}\left( \mathbf{s}+\mathbf{z} \right)=\mathbf{s},
\end{equation}
where $\mathbf{A}$ is a $(K+1) \times (K+1)$ matrix with the element $a_{kk'}$ of the real value for $k,k'= 0,1,\cdots,K$.  By assuming $\hat a^{(k)}_{p} = \sum\limits^{M^{(2)}_k}_{m =M^{(1)}_k} {{f_{\mu p - m}}{g_m}}$, the symmetry of matrix $\mathbf{A}$ can be found by 
\begin{equation} \label {3-9}
\begin{split}
a_{kk'} - a_{k'k} &= \hat a^{(k)}_{k-k'} - \hat a^{(k')}_{k'-k} \\
&= \sum\limits^{M^{(2)}_k}_{m =M^{(1)}_k} {{f_{\mu \left( {k - k'} \right) - m}}{g_m}}  - \sum\limits^{M^{(2)}_{k'}}_{m =M^{(1)}_{k'}} {{f_{\mu \left( {k' - k} \right) - m}}{g_m}}\\
&= 0.
\end{split}
\end{equation}

All elements of $\mathbf{A}$ are zero apart from $a_{kk'} = \hat a^{(k)}_{\left| {k-k'} \right|}$ for $k-k'=P_k^{(1)},P_k^{(1)}+1,\cdots,P_k^{(2)}$.  Then, a closed-form solution is obtained
\begin{equation} \label {3-10}
\mathbf{z} = \mathbf{A}^{-1}\mathbf{s} - \mathbf{s}.
\end{equation}

Since \eqref{3-10} involves the calculation of $\mathbf{A}^{-1}$ that requires the calculation complexity at $\mathcal{O}\left((K+1)^3\right)$, where $K+1$ is the length of the transmit signal block, the authors are making the simplification from decomposing $\mathbf{A}$, first, into $\mathbf{A} = \mathbf{A}_t - \mathbf{A}_p$, where $\mathbf{A}_t$ is found to be a $(K+1)\times(K+1)$ banded Toeplitz matrix generated by the vector $\hat {\mathbf{a}} = \left[ a^{(k)}_0,  a^{(k)}_1, \cdots,  a^{(k)}_{2\tau_0} \right]^{\dag}$ \cite {ref_Trench}. Then, $\mathbf{A}_p$ can be written as 
\begin{equation} \label {3-11}
\mathbf{A}_p = \left[
\begin{array}{*{20}{c}}
\mathbf{c}&\mathbf{0}&\mathbf{0}\\
\mathbf{0}&\mathbf{0}&\mathbf{0}\\
\mathbf{0}&\mathbf{0}&\mathbf{e}
\end{array} \right]
\end{equation}
with
\begin{equation} \label {3-12}
\setlength\arraycolsep{0pt}
\mathbf{c} = \left[ {\begin{array}{*{20}{c}}
{{\hat a_0} - \hat a_0^{(0)}}&{{\hat a_1} - \hat a_1^{(0)}}& \cdots &{{\hat a_{2{\tau _0}}} - \hat a_{2{\tau _0}}^{(0)}}\\
{{\hat a_1} - \hat a_1^{(0)}}&{{\hat a_0} - \hat a_0^{(1)}}&{}& \vdots \\
\vdots &{}& \ddots &{{\hat a_1} - \hat a_1^{(2{\tau _0} - 1)}}\\
{{\hat a_{2{\tau _0}}} - \hat a_{2{\tau _0}}^{(0)}}& \cdots &{{\hat a_1} - \hat a_1^{(2{\tau _0} - 1)}}&{{\hat a_0} - \hat a_0^{(2{\tau _0})}}
\end{array}} \right]
\end{equation}
and
\begin{equation} \label {3-13}
\setlength\arraycolsep{-1.5pt}
\mathbf{e} = \left[ {\begin{array}{*{20}{c}}
{{\hat a_0} - \hat a_0^{(K - 2{\tau _0})}}&{{\hat a_1} - \hat a_1^{(K - 2{\tau _0}+1)}}& \cdots &{{\hat a_{2{\tau _0}}} - \hat a_{2{\tau _0}}^{(K)}}\\
{{\hat a_1} - \hat a_1^{(K - 2{\tau _0}+1)}}&{{\hat a_0} - \hat a_0^{(K - 2{\tau _0}+1)}}& {} & \vdots \\
\vdots & {} & \ddots &{{\hat a_1} - \hat a_1^{(K)}}\\
{{\hat a_{2{\tau _0}}} - \hat a_{2{\tau _0}}^{(K)}}& \cdots &{{\hat a_1} - \hat a_1^{(K)}}&{{\hat a_0} -\hat  a_0^{(K)}}
\end{array}} \right].
\end{equation}
$\mathbf c$ and $\mathbf e$ are the two submatrices of $\mathbf{A}_p$ with the same dimensions of $(2\tau_0+1)\times(2\tau_0+1)$.  Actually, \eqref{3-11} deals with the $4\tau_0+2$ boundary terms at the output of the matched filter.  The signal block and AF set in \eqref{3-10} can be resolved to
\begin{equation} \label {3-14}
\mathbf{s} + \mathbf{z}= \mathbf{B}\mathbf{s}+\mathbf{B}\mathbf{A}_p(\mathbf{s}+\mathbf{z}),
\end{equation}
where $\mathbf{B}$ is the inverse of $\mathbf{A}_t$.

Further, we define $\mathbf{u}$ and $\mathbf{v}$ as the results of $\mathbf{c}\left(\mathbf{z}+\mathbf{s}\right)^h$ and $\mathbf{e}\left(\mathbf{z}+\mathbf{s}\right)^t$ in \eqref{3-14}, respectively. By using superscripts of $\left(\cdot\right)^h$ and $\left(\cdot\right)^t$ to the $(2\tau_0 + 1)\times 1$ vectors that contain the $2\tau_0+1$ head and terminal terms, the matrix $\mathbf{B}$ can be partitioned into nine blocks as
\vspace{-0.3cm}
\begin{figure}[h]
\centering
\includegraphics[width=\columnwidth]{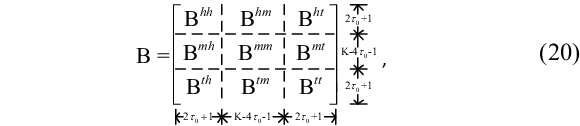}
\end{figure} \vspace{-0.6cm}
$\\ \\ $
where the superscripts of $(\cdot)^{hh}$, $(\cdot)^{mm}$, and $(\cdot)^{tt}$ indicate the head, middle, and terminal blocks, respectively.  While, superscripts of $(\cdot)^{hm}$ and $(\cdot)^{mh}$ denote the coupling terms.

Substituting the partitioned $\mathbf{B}$ into \eqref {3-14} yields
\begin{equation} \label {3-15}
\setcounter{equation}{21}
\setlength\arraycolsep{0pt}
\left[ {\begin{array}{*{20}{c}}
\left({{\mathbf{z}} + {\mathbf{s}}}\right)^h\\
\left({{\mathbf{z}} + {\mathbf{s}}}\right)^t
\end{array}} \right] = \left[ {\begin{array}{*{20}{c}}
{{{\left( {\mathbf{B}\mathbf{s}} \right)}^h}}\\
{{{\left( {\mathbf{B}\mathbf{s}} \right)}^t}}
\end{array}} \right] + \left[ {\begin{array}{*{20}{c}}
{{\mathbf{B}^{hh}}\mathbf{u} + {\mathbf{B}^{tt}}\mathbf{v}}\\
{{\mathbf{B}^{th}}\mathbf{u} + {\mathbf{B}^{tt}}\mathbf{v}}
\end{array}} \right] = \left[ {\begin{array}{*{20}{c}}
{{\mathbf{c}^{ - 1}}\mathbf{u}}\\
{{\mathbf{e}^{ - 1}}\mathbf{v}}
\end{array}} \right],
\end{equation}
which can be further simplified to
\begin{equation} \label {3-16}
\setlength\arraycolsep{2pt}
\left[{\begin{array}{*{20}{c}}
\mathbf{u}\\
\mathbf{v}
\end{array}}\right] = \mathbf{J}^{-1} \left[ {\begin{array}{*{20}{c}}
{{\mathbf{B}^{hh}}}&{{\mathbf{B}^{ht}}}\\
{{\mathbf{B}^{th}}}&{{\mathbf{B}^{tt}}}
\end{array}} \right]\left[ {\begin{array}{*{20}{c}}
{{\mathbf{s}^h}}\\
{{\mathbf{s}^t}}
\end{array}} \right],
\end{equation}
where
\begin{equation} \label{3-17}
\setlength\arraycolsep{2.5pt}
\mathbf{J} = \left[ {\begin{array}{*{20}{c}}
{{\mathbf{c}^{ - 1}}}&\mathbf{0}\\
\mathbf{0}&\mathbf{e}^{ - 1}
\end{array}} \right] - \left[ {\begin{array}{*{20}{c}}
{{\mathbf{B}^{hh}}}&{{\mathbf{B}^{ht}}}\\
{{\mathbf{B}^{th}}}&{{\mathbf{B}^{tt}}}
\end{array}} \right].
\end{equation}

Finally, the simplified solution is obtained by
\begin{equation} \label {3-18}
\mathbf{z} = \mathbf{B}\mathbf{s}-\mathbf{s}+\mathbf{B}
\left[ {\begin{array}{*{20}{c}}
\mathbf{u}\\
\mathbf{0}\\
\mathbf{v}
\end{array}} \right],
\end{equation}
where $\mathbf{B}$ is a non-singular matrix when the value of the Fourier transform result of its circulant vector is not zero.

\section{Performance Evaluation}
\begin{figure}
\centering
\vspace{-0.35cm}
\subfigure[]{
\centering
\includegraphics[width=0.45\columnwidth]{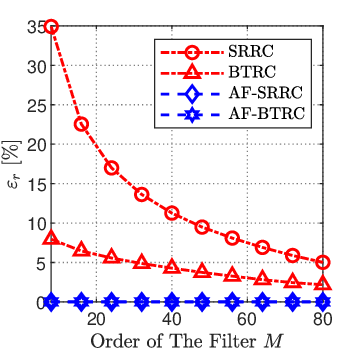}
}
\hspace{-0.5cm}
\subfigure[]{
\centering
\includegraphics[width=0.45\columnwidth]{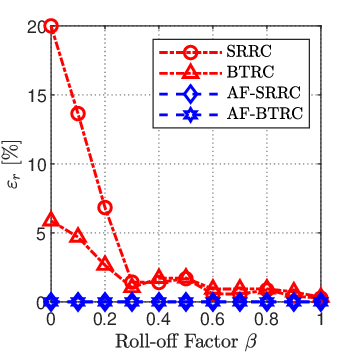}
}
\caption{Relative RMS error versus (a) order of the filter ($\mu = 4$, $\beta = 0.05$) and (b) roll-off factor ($\mu = 4$, $M = 24$).}
\label{fig_3}
\end{figure}

\begin{figure}
\centering
\vspace{-0.35cm}
\includegraphics[width=0.82\columnwidth]{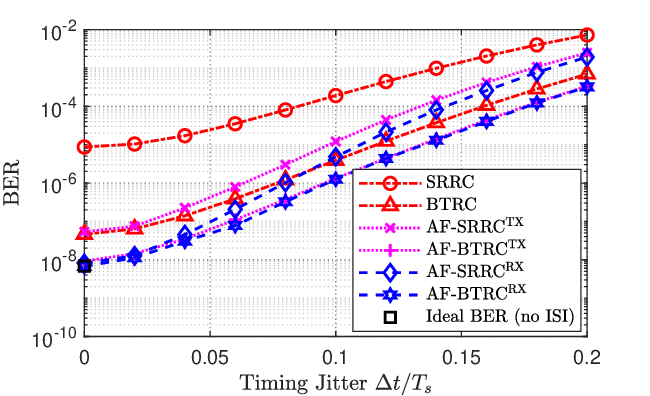}
\caption{BER versus timing jitter ($\mu = 4$, $M = 24$, $\beta = 0.05$, $E_b/N_0=12$ dB).}
\label{fig_4}
\end{figure}

In this section, the AF method is used to combine the conventional Nyquist filters SRRC \cite{ref_Proakis} and BTRC \cite{ref_Beaulieu}, which use the upsampling ratio $\mu = 4$ to form two new solutions referred to as the AF-SRRC and AF-BTRC, respectively.  The simulations are performed for comparisons with the conventional Nyquist filters.

To illustrate the zero-ISI performance of the AF-based methods, the comparisons are made upon the relative root-mean-square (RMS) error \cite{ref_Golub} defined as
\begin{equation} \label {4-1}
\varepsilon_r = \frac{{\left\| {\mathbf{y}  - \mathbf{s}} \right\|}}{{\left\| {\mathbf{s}} \right\|}} \times 100\%,
\end{equation}
where $\mathbf{s}$ and $\mathbf{y}$ are the transmit and receive symbols, respectively.

It is noted that the mathematical error $\varepsilon_r$ of the conventional Nyquist filters in \eqref{4-1} equals the RMS-power of the ISI, which is removed by the AF method in \eqref{2-5}.

The simulations are carried out by setting the timing jitter at $\Delta t/T_s=0$ and the roll-off factor at $\beta = 0.05$, where $\Delta t$ and $T_s$ are the time deviation from the zero-ISI point (promised by Nyquist condition) and the symbol duration, respectively. Figure \ref{fig_3}(a) compares $\varepsilon_r$ versus $M$ between the AF-based methods and the conventional ones.  It is found that the relative RMS errors of the former are zero in the full range of $M$ because of the zero-ISI, while the latter suffer from the errors that are reduced with an increase of $M$.  Figure \ref{fig_3}(b)  investigates $\varepsilon_r$ versus $\beta$ and shows that the AF-based methods can remove the relative RMS errors completely, while the two conventional Nyquist filters suffer obviously from the errors that are also reduced with an increase of $\beta$.

In fact, the reduction of $\varepsilon_r$ in Fig. \ref{fig_3} indicates the convergence of the Fourier expansion in the conventional methods.

\begin{figure}
\centering
\vspace{-0.35cm}
\subfigure[SRRC]{
\centering
\includegraphics[width=0.45\columnwidth]{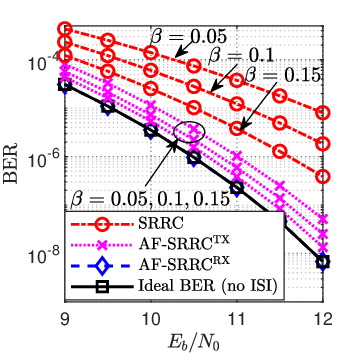}
}
\hspace{-0.5cm}
\subfigure[BTRC]{
\centering
\includegraphics[width=0.45\columnwidth]{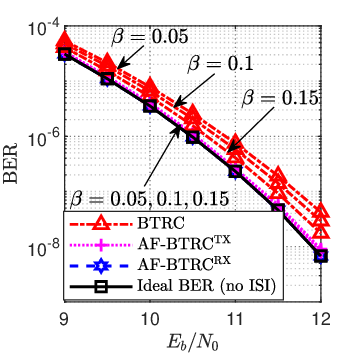}
}
\caption{BER versus $E_b/N_0$ ($\mu = 4$, $M = 24$) of BPSK for different roll-off factors.}
\label{fig_5}
\end{figure}

\begin{figure}
\centering
\vspace{-0.35cm}
\subfigure[SRRC]{
\centering
\includegraphics[width=0.45\columnwidth]{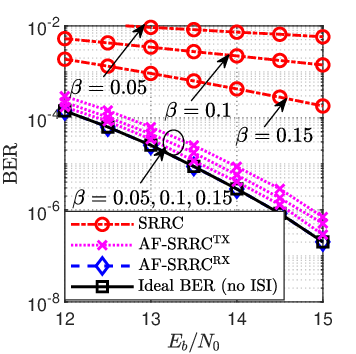}
}
\hspace{-0.5cm}
\subfigure[BTRC]{
\centering
\includegraphics[width=0.45\columnwidth]{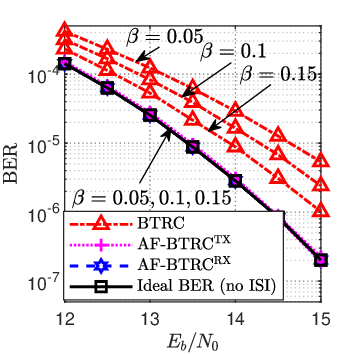}
}
\caption{BER versus $E_b/N_0$ ($\mu = 4$, $M = 24$) of 16-QAM for different roll-off factors.}
\label{fig_6}
\end{figure}

\begin{table*}[h]
\centering
\vspace{-0.35cm}
\caption{Power Comparisons for Various Roll-Off Factors by Setting $\mu = 4$, $M = 24$, and $\Delta t/T_s = 0$}
\label{tab_1}
\begin{tabular}{m{0.8cm}<{\centering}||m{0.8cm}<{\centering}m{0.8cm}<{\centering}m{1cm}<{\centering}m{0.8cm}<{\centering}m{1.2cm}<{\centering}m{1.2cm}<{\centering}||m{0.8cm}<{\centering}m{0.8cm}<{\centering}m{1cm}<{\centering}m{0.8cm}<{\centering}m{1.2cm}<{\centering}m{1.2cm}<{\centering}} \hline \vspace{0.1cm}
\vspace{0.1cm} & \multicolumn{6}{c||}{BPSK} & \multicolumn{6}{c}{16-QAM} \\ \cline{2-13}
\multicolumn{1}{c||}{$\beta$} & \vspace{0.1cm} The Nyquist Filters & \vspace{0.1cm} PAPR [dB] & \vspace{0.1cm} The AF Methods & \vspace{0.1cm} PAPR [dB] & \vspace{0.1cm} PAPR Diff. $[\%]$ & \vspace{0.1cm} $E_b^{\text{TX}}/E_b^{\text{RX}}$ [dB] & \vspace{0.1cm} The Nyquist Filters & \vspace{0.1cm} PAPR [dB] & \vspace{0.1cm} The AF Methods & \vspace{0.1cm} PAPR [dB] & \vspace{0.1cm} PAPR Diff. $[\%]$ & \vspace{0.1cm} $E_b^{\text{TX}}/E_b^{\text{RX}}$ [dB] \\ \hline \hline
0.05 & & \vspace{0.1cm} 6.0816 & & \vspace{0.1cm} 6.1027  & \vspace{0.1cm} 0.49 & 0.524  & \vspace{0.1cm} & \vspace{0.1cm} 8.7633 & & \vspace{0.1cm} 9.2989 & \vspace{0.1cm} 13.13 & 0.498 \\
0.1 & \multicolumn{1}{c}{SRRC} & 6.0574 & AF-SRRC & 5.9969 & 1.38 & 0.322 & \multicolumn{1}{c}{SRRC} & 8.7300 & AF-SRRC & 9.1468 & 10.07 & 0.287 \\
0.15 & & 5.9692 & & 5.8724 & 2.20 & 0.175 & & 8.7016 & & 9.0307 & 9.12 & 0.146 \\ \hline\vspace{0.1cm}
0.05 & & \vspace{0.1cm} 5.1375 && \vspace{0.1cm} 4.9037 & \vspace{0.1cm} 5.24 & \vspace{0.1cm} 0.046 & & \vspace{0.1cm} 7.7707 && \vspace{0.1cm} 7.6599 & \vspace{0.1cm} 2.52 & \vspace{0.1cm} 0.038 \\
0.1 & \multicolumn{1}{c}{BTRC} & 5.2387 & AF-BTRC & 5.0462 & 4.34 & 0.034 & \multicolumn{1}{c}{BTRC} & 7.9034 & AF-BTRC & 7.8169 & 1.97 & 0.028 \\
0.15 & & 5.4318 & & 5.3104 & 2.76 & 0.018 & & 8.1475 & & 8.1405 & 0.16 & 0.017 \\ \hline \hline
\end{tabular}
\end{table*}

In addition, the bit error ratios (BERs) are investigated for showing the advantages of the AF method.  The simulations of the AF-based methods are carried out using the two kinds of SNRs separately; one is estimated at the receiver and the other from the transmitter for the following reasons.  According to the digital communication model over the AWGN channel in Shannon's theory, the SNR is the ratio of the average energy of the received signal samples to that of the noise.  However, in this study, the Nyquist filter should be regarded as two channels that cascade the AWGN channel.  Hence, the input power of the AWGN channel, i.e. the transmit power, should be added to the simulations.

At first, we use the SNR measured at the receiver to test the BER performance of Binary Phase Shift Keying Modulation (BPSK) with respect to the timing jitter.  The BER is simulated by using the hard decision method at $E_b/N_0 = 12$, $M = 24$, and $\beta = 0.05$, where $M$ and $\beta$ are the filters' order and the roll-off factor, respectively.  The results are shown in Fig. \ref{fig_4}, where the subscript $(\cdot)^{\text{RX}}$ marks the case of the received SNR.  It is found that the AF method brings the BERs' reduction to the two conventional Nyquist filters significantly and leads to a lower sensitivity to the timing jitter simultaneously.

More BER simulations are carried out for testing the performances of 16-Quadrature Amplitude Modulation (QAM) at $\Delta t/T_s = 0$.  The results are plotted in Figs. \ref{fig_5} and \ref{fig_6}, where the improvements of the AF method can be clearly found again.  Actually, the BER performances of the AF-based methods are exactly the same as the ideal results (without any filters) because of the zero-ISI performance.

Secondly, the average energy of the signal samples at the transmitter in \eqref{2-1} is used to calculate the SNR.  The simulations are performed in parallel to the above procedures.  The BER results are plotted with the mark of $(\cdot)^{\text{TX}}$ in Figs. \ref{fig_4}-\ref{fig_6}, where some BER degradations are found.  The quantitative comparison of the SNRs can be made by
\begin{equation}\label{4-2}
E_b^{\text{TX}}/E_b^{\text{RX}} = \frac{\left\| \mathbf{s}'\right\|^2}{\left\| \mathbf{s} \right\|^2}
\end{equation}
where $E_b^{\text{TX}}$ and $E_b^{\text{RX}}$ are the bit energy accounted at the signal samples of the transmitter and receiver, respectively, and $\mathbf{s}'$ and $\mathbf{s}$ are the transmit and receive signals' samples in \eqref{2-1}.  The results of \eqref{4-2} are listed in Tab. \ref{tab_1} for BPSK and 16-QAM as well.

Finally, the problem of peak-to-average power ratio (PAPR) is investigated by using $M = 24$ and $\Delta t/T_s = 0$.  The results of SRRC, AF-SRRC, BTRC, and AF-BTRC are listed in Tab. \ref{tab_1}, which show the non-significant impact of the AF method on the PAPR performance since the largest difference is found smaller than 6\% for BPSK.  Even for 16-QAM, the largest difference is found smaller than 14\%.

Before ending this paper, it is noted that the communication bandwidth of the transmission does not change at all, since the AFs are actually a part of the input signals to the filters.  While, two drawbacks of the AF method are analysed for the delay and additional complexity problems in the following two paragraphs.

The delay can be attributed to the requirement of taking all transmit signals for the calculation of the AFs.  Hence, the delay time is, at least, equal to the transmission time of the whole signal block.

The implementation complexity is analysed from \eqref{3-18}, where the matrix $\mathbf{B}$ has been decomposed in terms of the circular decomposition \cite{ref_Jain} leading to the calculation complexity at $\mathcal{O}\left( (K+1) {\rm log}_2 (K+1)\right)$.  While the inverse $\mathbf{y}^{-1}$, $\mathbf{e}^{-1}$ and $\mathbf{J}^{-1}$ can be calculated by the Trench algorithm \cite {ref_Trench} at $\mathcal{O}\left((M+1)^2 \right)$.   Since, in most communication operations, the length of transmit signal block is much longer than the Nyquist filter in terms of the coefficients' number, the complexity can be dominated by $\mathcal{O}\left( (K+1) {\rm log}_2 (K+1)\right)$.

\section {Conclusion}
The AF method is introduced to compensate shortcomings of the truncated Fourier transform in the conventional Nyquist filters.  In fact, the AF method can be applied generally to all the Nyquist filters for removing the ISI, though the theoretical derivations of this paper are given only to the paired filters.  Actually, the merit of the AF method lies in the fact that the zero-ISI can be realised regardless of the order of the Nyquist filters.  In addition, the lower sensitivity to the timing jitter is also found in comparison with the conventional Nyquist filters.  Finally, the delay and complexity problems are analysed for the application of the AF method.

\section*{Acknowledgement}
The authors would like to thank the editor, Dr. Ferdi Kara, and the anonymous reviewers for their valuable comments to improve the presentation of this paper.

\balance

\end{document}